# UML Modeling to TM Modeling and Back

Sabah Al-Fedaghi
*sabah.alfedaghi@ku.edu.kw*
Computer Engineering Department, Kuwait University, Kuwait

**Summary**
Certainly, the success of the Unified Modeling Language (UML) as the de facto standard for modeling software systems does not imply closing the door on scientific exploration or experimentation with modeling in the field. Continuing studies in this area can produce theoretical results that strengthen UML as the leading modeling language. Recently, a new modeling technique has been proposed called thinging machine (TM) modeling. This paper utilizes TM to further understand UML, with two objectives:
(a) Fine issues in UML are studied, including theoretical notions such as events, objects, actions, activities, etc. Specifically, TM can be used to solve problems related to internal cross-diagram integration.
(b) TM applies a different method of conceptualization, including building a model on one-category ontology in contrast to the object-oriented paradigm. The long-term objective of this study is to explore the possibility of TM complementing certain aspects in the UML methodology to develop and design software systems.
Accordingly, we alternate between UML and TM modeling. A sample UML model is redesigned in TM, and then UML diagrams are extracted from TM. The results clarify many notions in both models. Particularly, the TM behavioral specification seems to be applicable in UML.

## 1. Introduction

Exploring modeling in the Unified Modeling Language (UML) context is important for progress in conceptual modeling in software engineering. According to [1], before UML standards, diagrammatic software modeling was plagued by the incompatibility of different notations, the absence of standardized notation, and the tiny and fragmented nature of the modeling tools market. Of the few tools that were available, many only allowed sketching of software designs and design documentation, but were rarely integrated into the software development life cycle. Now, UML has become the lingua franca of software development, supported by every major commercial IT vendor as well as a flourishing selection of open-source tools.

### 1.1 UML Advantages and Disadvantages

Most software professionals are at least acquainted with, if not well-versed in, UML diagrams, making it the go-to option to explain software design models [2]. According to [2], "What makes UML well-suited to and much-needed for software development is its flexibility. UML is a rich and extensive language that can be used to model not just object-oriented software engineering, but application structure, behavior, and business processes too." For [3], documentation and modeling are perhaps two of the most difficult tasks within the software development process. The absence of design documentation is fine in the short term, but it can become a problem in the long run, and UML has become a huge help in such circumstances to alleviate ambiguity and questions regarding the design [2]. According to [2], there is no holistic or appropriate substitute for UML. Domain-specific languages for diagrammatic modeling have been introduced, but none of them has found wide acceptance, supporting UML as the best option among diagrammatic languages [2]. For some researchers, UML has become synonymous with software modeling [4]. The use of UML as a language leads to an improvement in cooperation between technical and nontechnical competencies. It helps in better understanding systems, in revealing simplification, and in easier recognition of possible risks. Through early detection of errors, costs can be reduced during the implementation phase [5].

On the other hand, a considerable portion of software developers do not use UML [2]. There is no need for a UML diagram to communicate designs. One can do that with informal box-and-line diagrams, such as those drawn in PowerPoint. Additionally, UML has grown in complexity, which makes many people feel as though they are better off without it [2]. Complexity is the number one problem in software, and according to [6], "our co-workers don't see the complexity that they have created and no amount of knowledge or expertise will slay essential complexity."

An important motivation for this paper is that UML has become part of many software engineering course curricula at universities worldwide. It is common for students to have difficulty absorbing UML due to its complexity [7]. Often, students think that UML diagrams are useless and serve only as documentation that no one reads [8].

The number of UML diagram types is still a disturbing issue. The model multiplicity problem [9] concerns the integrated view of structure and behavior. The issue concerns how UML diagrams are associated with one another. Génova and Nubiol [3] considered such a matter to be primarily related to the role of software development methodology, which can best answer this question. Soffer [10] argued that because UML evolved bottom-up from object-oriented programming concepts, it lacks a system-theoretical ontological foundation encompassing observations about common features characterizing systems regardless of domain. In this context, the Object-Process Language (OPL) [11-12] was proposed as a modeling language that is both formal and intuitive. OPM was developed as a comprehensive approach to systems engineering that integrates function, structure, and behavior in a single unifying model [11]. OPM is specified as ISO/PAS 19450 [13], and can be considered an alternative to UML. Applying thinging machine (TM) modeling in the context of OPM is as important as applying it to UML.

1.2 Motivation and Objectives

Certainly, the success of UML as the de facto modeling language in software engineering does not imply ceasing scientific exploration or experimentation with other methodologies. Continuing studies in this area can produce theoretical results that strengthen UML as the leading modeling language. Recently, a new modeling technique has been proposed, called TM modeling [13-18]. TM is a good tool to analyze different modeling techniques, such as data flow diagrams, flowcharts, etc. This paper connects TM to UML with two objectives:

(a) TM is used to handle fine issues in UML by investigating several theoretical notions such as events, objects, actions, activities, etc. Specifically, TM can be used to investigate UML problems related to internal cross-diagram integration.

(b) TM applies a different method of conceptualization, including building a model on one-category ontology in contrast to the object-oriented paradigm. The long-term objective of this study is to explore how to align TM with the UML methodology to develop and design software systems.

Accordingly, we alternate between UML and TM modeling. A sample UML model is redesigned in TM, and then UML diagrams are extracted from the TM model. Our results clarify many notions in both models. Particularly, the TM behavioral specification seems to be applicable in UML.

1.3 Outline

This paper is organized as follows:
1. Section 2 provides a brief overview of TM.
2. Section 3 includes a sample UML model from Chapter 5 of Ian Sommerville's book [20], which was developed as part of the requirements engineering and system design processes. According to Sommerville [20], after one has read the chapter, one will
   - Understand how diagrammatic models can be used to represent software systems and why several types of model are needed to fully represent a system;
   - Understand the fundamental system modeling perspectives of context, interaction, structure, and behavior;
   - Have been introduced to model-driven engineering, where an executable system is automatically generated from structural and behavioral models. [20].

   Sommerville's UML model involves developing the specification for a mental health care (Mentcare) patient information system. This system is intended to manage information about patients attending mental health clinics.
3. Sommerville starts the modeling process with an activity diagram for the Mentcare system; hence, we construct a corresponding TM model. Our strategy is to fit all UML diagrams into one TM diagram.
4. Next, Sommerville introduces use case diagrams for the Mentcare system; accordingly, we extend the TM model to assimilate these use cases.
5. Subsequently, Sommerville presents sequence diagrams. It is interesting to note that the contents of such diagrams are already modeled in the TM model in Step 4.

6. The UML diagram is also incorporated into the TM model through a sample association between two classes. The association is applied as a constraint diagram inside the extended TM diagram.
7. Lastly, Sommerville progresses to UML modeling of the behavior. Here, Sommerville turns to a state diagram of a microwave oven. It is not clear why the state diagram of the Mentcare system is not developed. However, we continue modeling the behavior of the Mentcare system by developing the TM events diagram and behavior diagram.

## 2. Thinging Machine Modeling

This section presents a summary from published papers [13-18] that briefly describes TM modeling. TM modeling seems to be a promising methodology and has been applied in many areas, such as network documentation, robot architecture, and security.

The term *thinging* in TM comes from Heidegger [21], in whose works thinging expresses how a "thing things," which he explained as gathering or tying together its constituents. In TM modeling, things are unified with the concept of a process by being viewed as single ontological things/machines, or *thimacs*, which populate the world. A unit in such a universe has a dual role as a thing and as a machine. *Machine* refers to the abstract machine shown in Fig. 1, which is a generalization of the known input-process-output model. Input and output are lumped together in the *transfer* stage (Fig. 1). This represents the machine's gate, where things flow to and from other machines. The *release* stage is a waiting stage for things in the machine until transfer is activated (e.g., goods produced by a factory are stored until transported in trucks). The *receive* stage represents the phase in which things flowing from other machines *arrive* to be *accepted* inside the machine or sent back outside (e.g., wrong address). For simplicity's sake, in the modeling examples in this paper, we assume that things that arrive are always accepted; hence, we always use the receive stage in these examples.

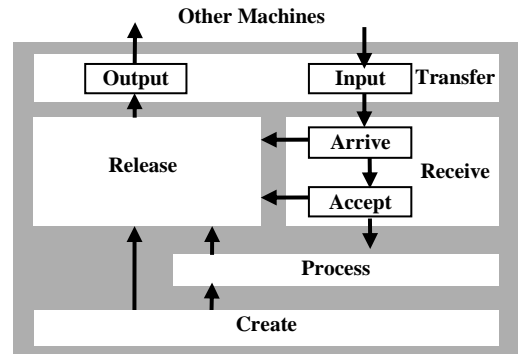

Fig. 1 Thinging machine model.

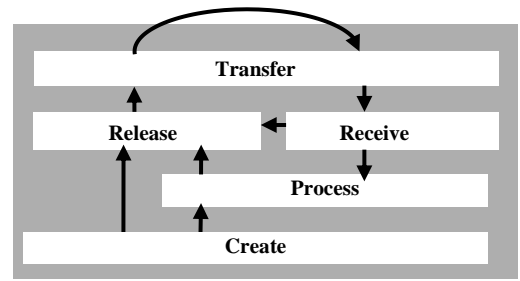

Fig. 2 Simplified TM model.

The *create* stage in Fig. 1 denotes the appearance of a new thing in the machine (e.g., a generator's output after converting a form of energy into electricity). The process stage in Fig. 1 refers to changing the form of a thing without generating a new entity (e.g., transforming a decimal number to binary form). A stage in TM might include a storage area (represented by a cylinder) that accommodates things inside the stage. Fig. 2 shows a simplification of Fig. 1, where things flow from the output of one machine to the input of another machine.

A thing in TM modeling is whatever is created, processed, released, transferred, or received. Thus, a machine creates, processes, releases, transfers, and receives things. Hence, create, process, release, transfer, and receive are called *actions*. The arrows in Fig. 1 denote the flow of things from one stage to another or within the machine.

The TM model is the grand machine that results from smaller machines. To facilitate shifting among flows (e.g., processing electricity creates cold air), the TM model includes triggering, denoted by dashed arrows.

## 3. From UML to TM Modeling

Note that the TM produces three models: static, event, and behavioral models. In UML, one possible approach to the order in which project diagrams can be implemented is use case diagram, activity diagram, class diagram, sequence diagram, and state diagram [3]. In Chapter 5 of [20], a sample UML model is developed as part of requirements engineering and system design processes. This order may be iterative—for example, according to [22], we can identify classes from dynamic models and public operations in classes from actions and activities in state chart diagrams, as well as activity lines in sequence diagrams. Sommerville's book *Software Engineering* [20] roughly follows these phases of development. Erickson and Siau [23] showed that five UML diagram types could represent most system essentials. Hence, [20] concentrates on these five UML diagram types: activity diagrams, use case diagrams, sequence diagrams, class diagrams, and state diagrams.

3.1 Context Models

According to [20], at the first stage in the specification of a system, we should decide on the system boundaries—that is, on what is and is not part of the system being developed. The Mentcare system, for example, is intended to manage information about patients attending mental health clinics. Simple context models are used, along with others such as business process models (pre-UML diagrams). Sommerville [20] gives a context model that shows the Mentcare system and the other systems in its environment.

3.2 Starting Point: Activity Diagram

Then, [20] starts with a UML activity diagram (Fig. 3) to show where the Mentcare system is used in an important mental health care process—involuntary detention. Sommerville [20] mentions that the UML activity model illustrates *how the software transforms an input to a sequence of commands*. There is no definition of what this activity is, besides examples such as the commands analyze, compute, and control. We can assume that all commands are activities. "Generic activities" are mentioned, but not defined. *Activity diagrams* show the activities involved in a process or in data processing. Note that in [20]'s discussion, such activities are different from the so-called process activities (e.g., collecting requirements). UML activity diagrams can be used to show the *business processes* in which systems are used. A UML activity diagram also shows *where* the Mentcare system is used in the mental health care process.

The result of such a discussion of the notion of activity is a vague idea of what an activity is. This is a common feature in UML literature. Many resources deal with what an activity diagram is, but rarely is the question of what defines an activity raised. An activity may be defined as a kind of operation of the system. Some modelers declare that an *activity* represents a *behavior* that is composed of individual elements, which are *actions* [24]. The activity diagram is based on an ambiguous notion. Sommerville's starting with the activity diagram to build the Mentcare system is not a promising beginning for modeling.

In contrast, the TM model is built on five generic actions, as specified in the previous section. However, we have no choice but to follow [20]'s UML path; we develop the TM that corresponds to the activity diagram of Fig. 3, as shown in Fig. 4.

- In Fig. 4, a person (circle 1) is brought to the Mentcare system (2). In the activity diagram, it is not clear who made such a decision; hence, we assume that such an action happens in the Mentcare facility. A detention decision is made (3) and communicated (4) to the person. Rights are also provided (5 and 6).
- If the person is dangerous (7) and there is no secure location available (8), then the involved person is sent to the police station (9). If the person is dangerous (10) and a secure location is available, the person is sent to that place (11). If the person is not dangerous (12), he or she is admitted as a patient (13).
- In the three cases of sending the person to the police station (9), sending the person to a secure location (11), and admittance to the hospital (13), information is generated (14) and sent to the social services (15), the next of kin (16), and the Mentcare information system (17, 18, and 19).

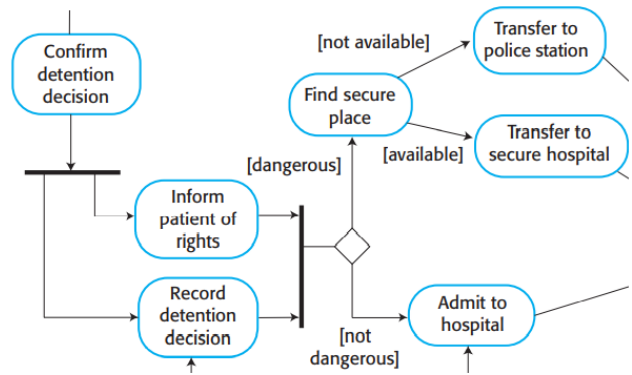

Fig. 3 Mentcare system activity diagram (partial from [20]).

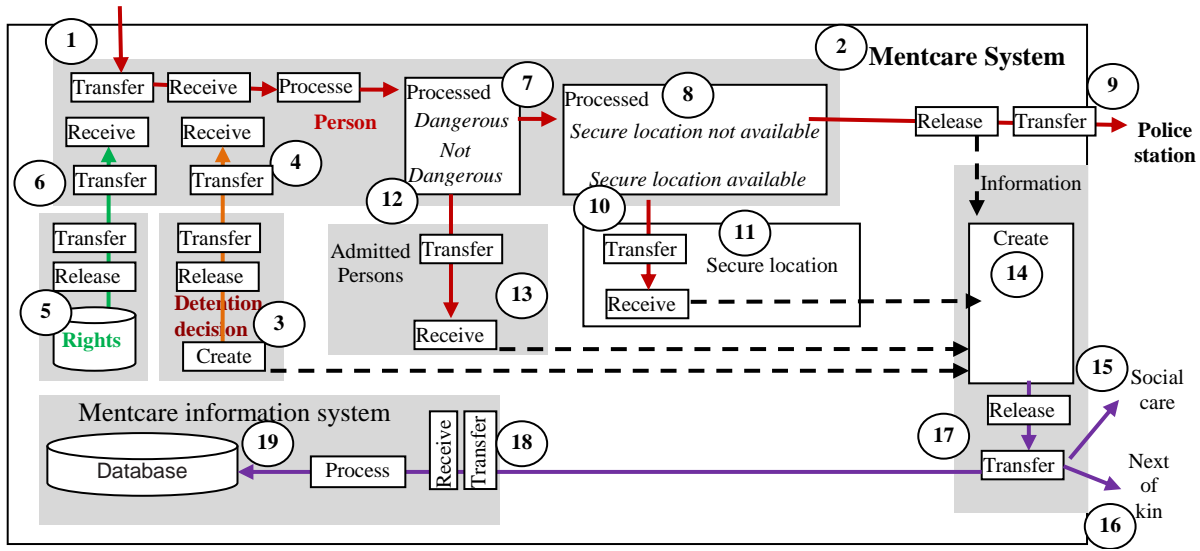

Fig. 4 TM model that corresponds to the activity diagram (see Fig. 3).

It can be observed that the TM representation is more complex than the activity diagram. However, if less precise description is desired, we can eliminate the release, transfer, and receive stages under the assumption that the direction of arrows is sufficient to express the flows in the model. With this simplification, Fig. 5 is produced. Additionally, it is not difficult to return to the activity diagram if desired. Fig. 6 shows an initial step in this direction.

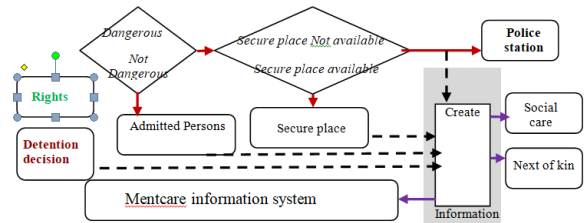

Fig. 6 A step toward converting TM diagram to activity diagram.

### 3.3 Step Two: Use Case Diagrams

Sommerville [20] continues modeling the Mentcare system by introducing three use case diagrams (see partial representation in Fig. 7):
- Medical receptionist/patient record system
- Tabular description of the transfer data use case
- A diagram that maps the medical receptionist registering a patient, unregistering a patient, viewing patient information, transferring data, and contacting the patient.

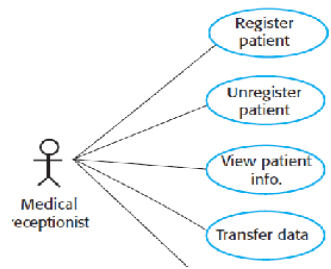

Fig. 7. Use case diagram (partial from [20]).

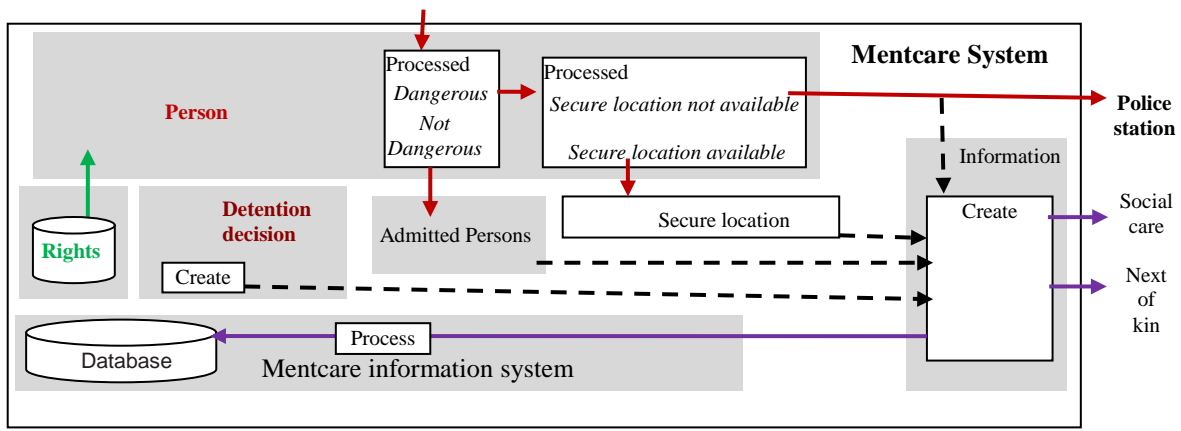

Fig. 5 Simplification of Fig. 4.

Our strategy is to include all UML diagrams in one TM diagram. Fig. 8 shows an extension of the TM model that corresponds to the activity diagram and includes descriptions in the use cases above.

We find that the medical receptionist (20) is the one in charge of the information system. The medical receptionist has a security clearance to access the system (21). Access is requested (22 and 23) and permission is given (24 and 25).

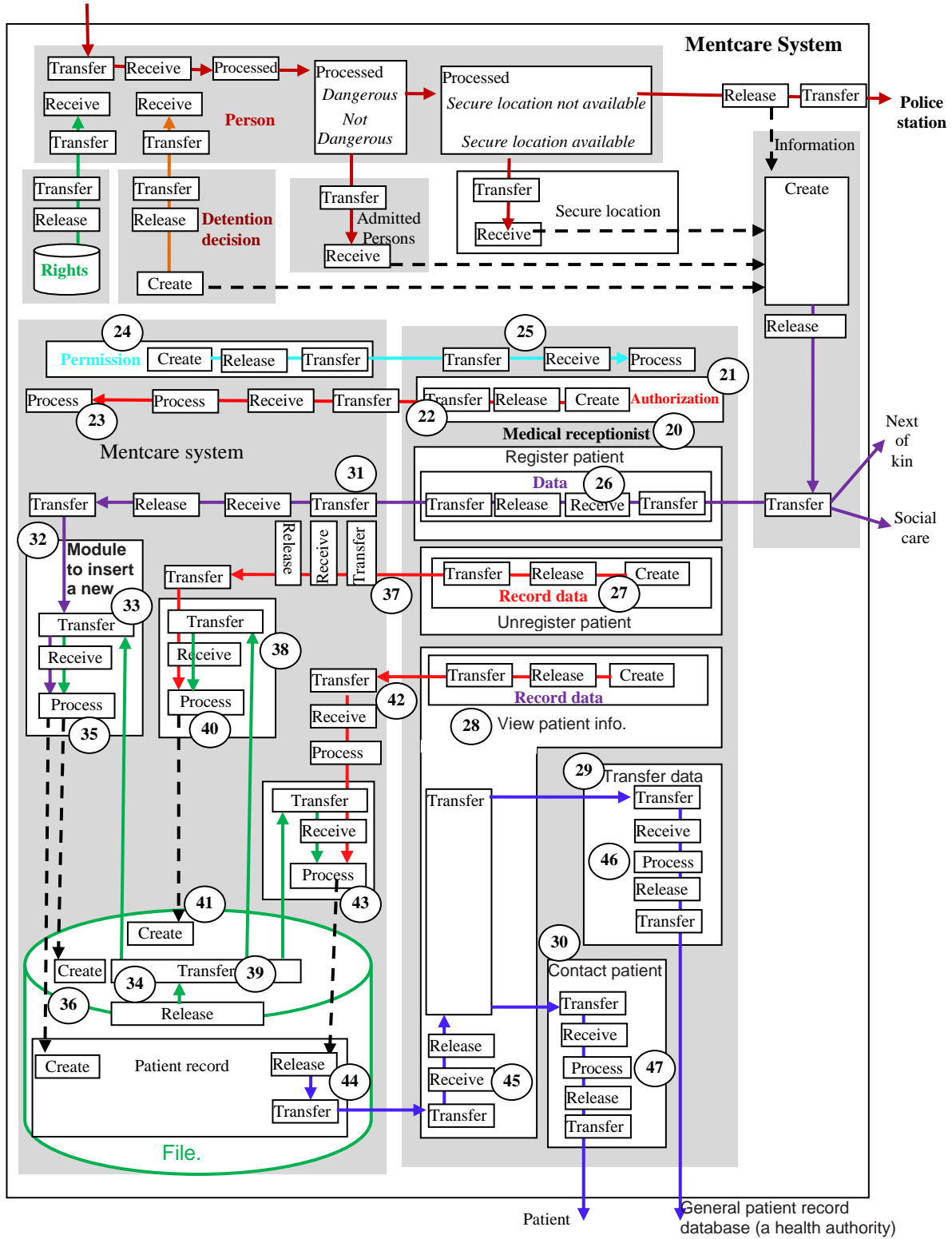

Fig. 8 TM model, extended to include use cases.

Additionally, the medical receptionist can register the patient (26), unregister the patient (27), view patient information (28), transfer data (29), and contact the patient (30).

**Register patient**: After receiving permission to access the information system, the medical receptionist inputs the patient data, which flows to the system (31). The module that handles constructing a new record takes this data (32a) along with (32b) the current patient file (33 and 34) and processes them (35). The result of this processing is a new version of the file that contains the new record. Note that the triggering (35 and 36) in the diagram hides the details of how to insert a record into a file.

**Unregister patient**: The medical receptionist sends an identifier of the record requiring deletion to the system (37). The module that facilitates this deletion takes (37a) this identifier along with (37b) the patient file (38 and 39) to process them (40), producing a new version of the file without the deleted record (41). Note that the triggering (40 and 41) hides the details of how to delete a record from a file.

**View patient info**: The medical receptionist inputs the record data (e.g., key) into the system (42). Processing a record key (43) along with the patient file triggers the retrieval of the required record (44) and sends it to the medical receptionist (45).

**Transfer data and contact patient**: These two functions involve processing (46 and 47) the received record (45), e.g., formulating the information and sending the resulting message to the patient or the general patient record database (a health authority).

### 3.4 Sequence Diagrams

Sommerville [20] continues modeling the Mentcare system by introducing two sequence diagrams. Usually, in UML, sequence diagrams are derived from use cases [22]. Sommerville's [20] first sequence diagram models the interactions involved in the "view patient information" use case, where a medical receptionist can see some patient information. This is already incorporated in the TM diagram of Fig. 8. It involves an extension of the authorization (Circle 21 in Fig. 8) to respond with an error message if the request for authorization is denied, which can easily be added to the TM diagram. The second sequence diagram is another example of a sequence diagram that illustrates additional features including direct communication between actors (e.g., medical receptionist and others).

### 3.5 Class Diagrams

Sommerville [20] continues modeling the Mentcare system by introducing class diagrams. First, [20] gives the association shown in Fig. 9. In TM modeling, such an association is viewed as a constraint. A constraint can be incorporated in the TM diagram in the usual way. The involved constraint can be viewed as meaning that when any record is created in the patient file, there should only be one record for any given patient. This is modeled in a new version of the static model (Fig. 10). When a record arrives to be registered in the patient's file (Circle 48), along with the file (49)—actually the file address—they are first processed by the constraint module (50 and 51). Note that the file flow to the constraint module is a conceptual flow (it may involve the address of the file). If the record is not in the file (51), then the record is inserted into the file to create a new version of the file (52 and 53).

Other descriptions given by [20] can easily be inserted in the TM diagram. For example, [20] introduced a consultation class, where each patient may be associated with several consultation records. In TM, we add the TM machine that an admitted patient goes to a doctor for consultation. In each meeting, a consultation record is created and sent to the medical receptionist, who inputs the consultation record after linking it to the patient record. We will not extend the static TM further because the process of combining all UML diagrams into a single TM diagram has become clear.

However, [20] does not give a clear description of the class structure in this example. The patient record is specified within the use case as "a receptionist may transfer personal information (address, phone number, etc.)." To illustrate the class notion within TM modeling, Fig. 11 shows a simplified TM specification of the flow of data from the receptionist screen to construct a record in the patient's file. Such TM specifications can be developed to any level of granularity.

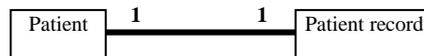

Fig. 9 Association between two classes (redrawn from [20]).

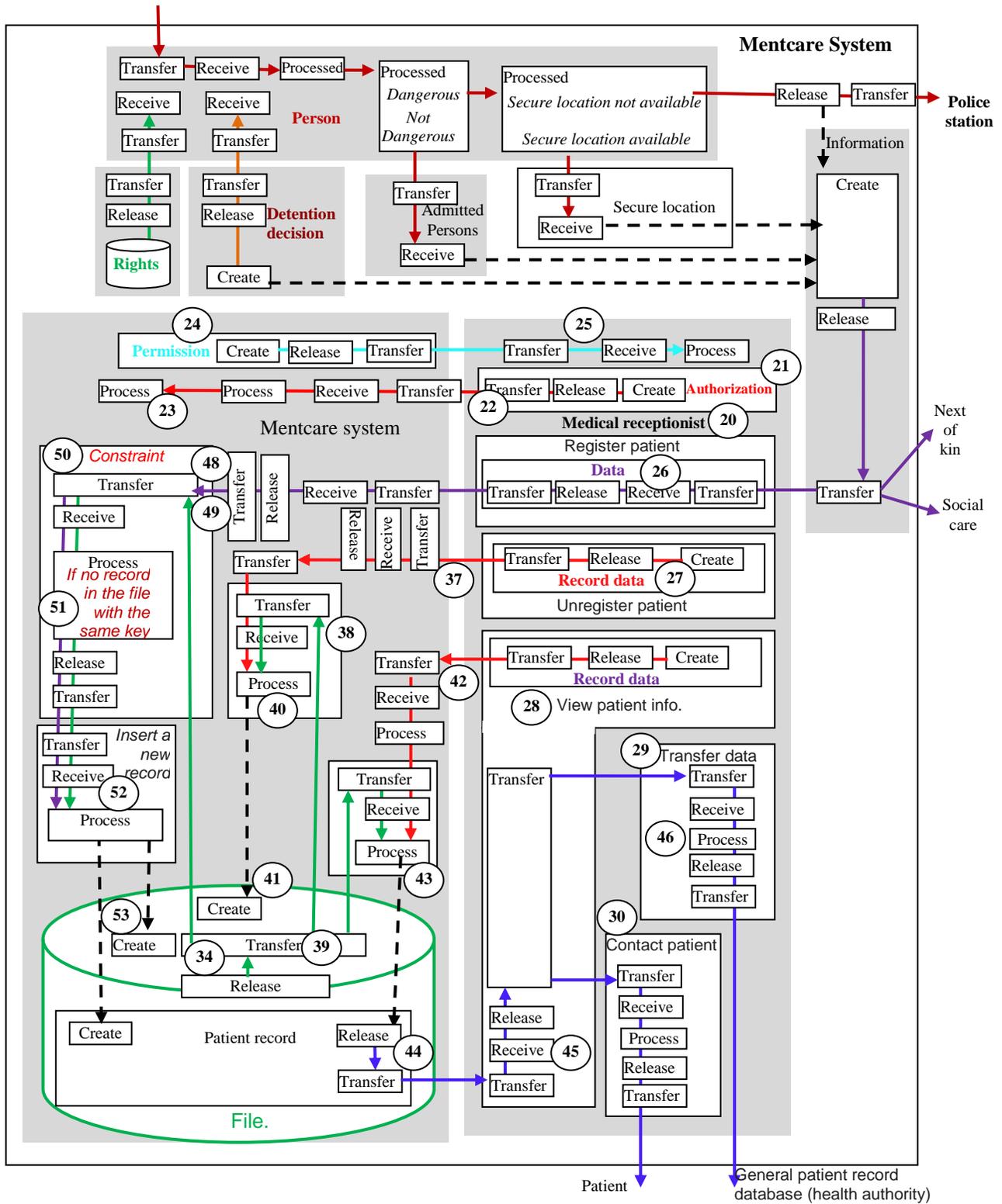

Fig. 10 TM model after including the constraint that each patient has only one record (circle 51).

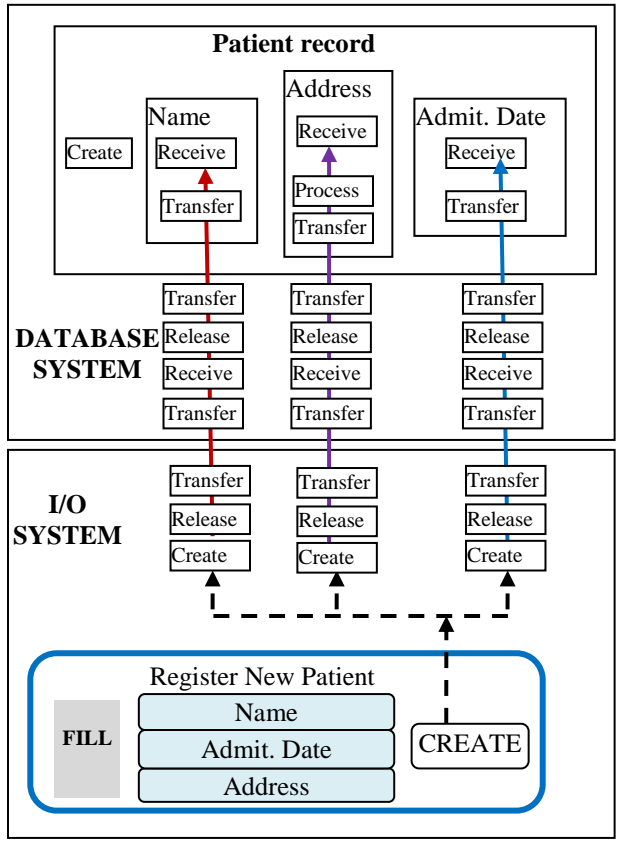

Fig. 11 Simplified TM specification of the notion of class and its methods.

3.6 Modeling the Behavior of the Mentcare System

Modeling of behaviors concerns the description of processes, chronological dependencies, state changes, the treatment of events, etc. In UML, behavior does not exist independently, but rather always affects certain objects. The execution of a behavior can always be traced to an object [3].

Sommerville [20] continues to model the behavior of the Mentcare system. The behavior mode is based on UML state diagrams, which show how the system reacts to internal and external events. State diagrams are the crucial notation to relate data aspects and behavior of objects. Usually, one state diagram is constructed for each class with important dynamic behavior [22]. Behavior models show what happens or what is supposed to happen when a system responds to a stimulus from its environment (e.g., data triggers processing, or an event happens that triggers system processing). Data flow diagrams that can be represented in UML are mentioned using the activity model of an insulin pump's operation. Activity diagrams are sometimes viewed as a special type of state diagram [19]. UML sequence diagrams of business processes are also discussed as an alternative way of showing the sequence of processing in a system.

Finally, state diagrams are discussed to support UML event-based modeling. State diagrams show system states and events that cause transitions from one state to another [20]. A state diagram is described here, but it is a state diagram of a microwave oven—there is no behavior model of the Mentcare system. Perhaps the connection to behavioral aspects is given by the methods of the class diagram, but this is not stated explicitly. This is typical of UML modeling materials: diagrams are presented in a fragmented way, and the discussion gives the behavioral description for a small example of completely different problems. Accordingly, we have to abandon the UML model of the Mentcare system and continue to behavioral modeling using only the TM model.

## 4. Modeling the Behavior of the Mentcare System in TM

We start by defining what an event is. An event is a machine that has
(a) a time submachine,
(b) a region submachine where the event happens, and
(c) another submachine (e.g., intensity). This characteristic will not be used in the Mentcare case study.

For example, in the context of the Mentcare model, Fig. 12 shows the model of the event *John was sent to the police station on 1/1/2021 because a secure location was not available*. The region is a sub-diagram of the static model. Note that such an event is not primitive (genetic). Genetic events are built upon the regions of genetic actions.

In UML, four different specifications are provided for behavioral descriptions:
• State diagrams
• Activities and actions
• Interactions
• Use cases [3]

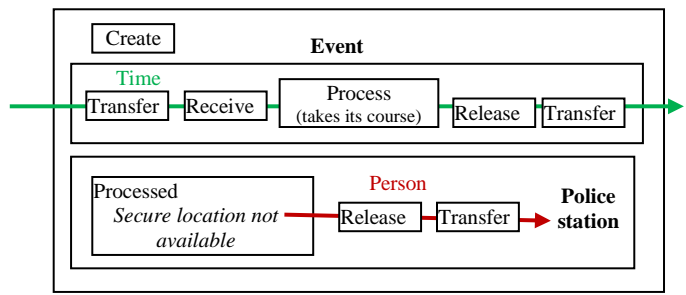

Fig. 12 The event *John was sent to the police station on 1/1/2021 because a secure location was not available*.

UML behavior is principally event oriented. The execution of behaviors is always triggered by an event. Behavior can be started either directly or indirectly via a trigger such as when the medical receptionist registers a patient, as modeled in a use case diagram. However, it is clear by now that in a use case, when the Mentcare receptionist is connected by an arrow to "register patient," this is not an event but a region, which may not be a region of event. It is a conceptual region, which may or may not be a space region. According to [3], a use case diagram is actually a structure diagram: it does not describe processes and behavioral patterns, but only relationships.

A TM static model can be constructed. For example, we can construct the static TM description for *Alice walked in the direction in which the March Hare was said to live.* The model is a TM region that has actions and a chronology of these actions but cannot be event-ized because it never happened. Similarly, the Mentcare receptionist connected by an arrow to "register patient" in the use case is a region of (possible) event when time is applied to the region. Hence, use case interactions, activities, and state diagrams in UML *do not involve behavioral modeling*.

Accordingly, we identify the following events in Fig. 13.
  Event 1 ($E_1$): A person is brought to the Mentcare center.
  Event 2 ($E_2$): A detention decision is made, and the person is informed.
  Event 3 ($E_3$): The detainee is informed of his or her rights.
  Event 4 ($E_4$): The detainee is examined and found to be dangerous.
  Event 5 ($E_5$): The detainee is transferred to the police because no other secure location is available.
  Event 6 ($E_6$): The detainee is transferred to a secure location that is available.
  Event 7 ($E_7$): It is determined that the person is not dangerous.
  Event 8 ($E_8$): Information about the detainee is sent to the social services, his or her next of kin, and a medical receptionist.
  Event 9 ($E_9$): The medical receptionist requests authorization to access the system and receives approval.
  Event 10 ($E_{10}$): The medical receptionist generates a request to register a patient, which flows to the information system.
  Event 11 ($E_{11}$): The medical receptionist generates a request to unregister a patient, which flows to the information system.
  Event 12 ($E_{12}$): The medical receptionist generates a request to view patient information.
  Event 13 ($E_{13}$): The information system checks whether a new patient is already in the system.
  Event 14 ($E_{14}$): The information system creates a record for the new patient in the database.
  Event 15 ($E_{15}$): The information system unregisters a patient.
  Event 16 ($E_{16}$): The information system retrieves the requested patient information.
  Event 17 ($E_{17}$): The requested patient information flows to the medical receptionist.
  Event 18 ($E_{18}$): The medical receptionist sends the information to the general patient record database (a health authority)
  Event 19 ($E_{19}$): The medical receptionist contacts the patient.
Fig. 14 shows the behavioral model of the Mentcare system.

## 5. Conclusion

This paper aimed at establishing connections between TM and UML, with a long-term objective of exploring the possibility of aligning a new modeling methodology, TM, with the UML. Additionally, we utilized TM as a tool to further understand UML. We remodeled a sample UML model in TM, then extracted some UML diagrams from TM.

Our results indicate that it is difficult to align the two representations. A TM can replace the activity diagram, use case, and sequence and class diagrams, considering the ontological differences such as semantic differences between object-oriented notation and TM elements (e.g., things). However, the issue still needs more investigation, especially in the context of system behavior. One result seems to indicate that TM can be used at a higher level of abstraction and thus provides a conceptual foundation for UML. This would make some UML diagrams obsolete. UML can take the idea of building the behavior of the system for the same program without introducing different notation. Similarly, in TM, the static diagram is divided into decompositions to insert time, thus facilitating the construction of events.

One clear conclusion of this work is that TM modeling can be used to understand the fundamental ideas and structures of other modeling techniques. With respect to UML, several future studies are needed because of its size and complexity. Future studies should focus on samples of UML models that have extensive behavioral representations, especially state diagrams that complement other diagrams over the same domain.

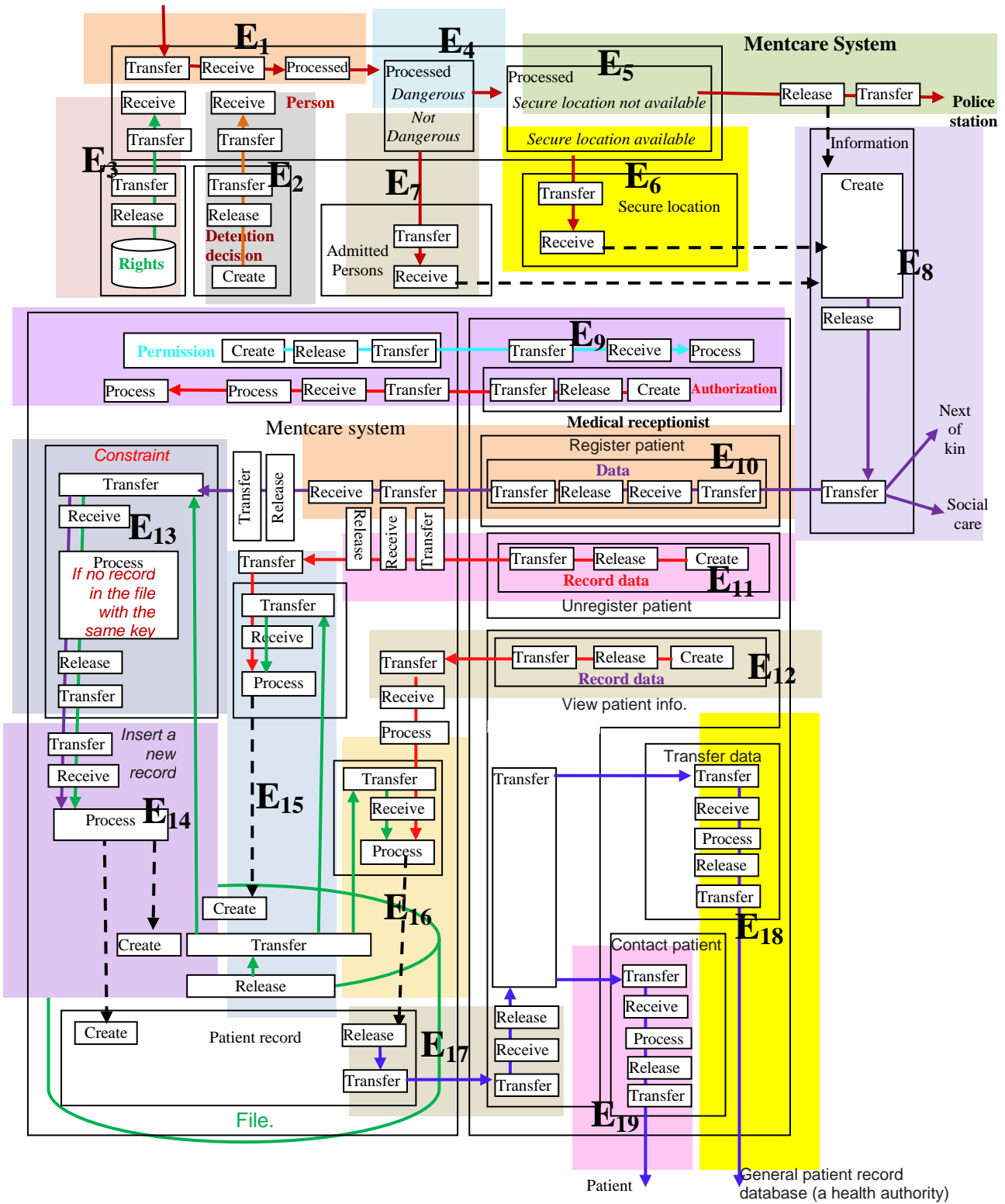

Fig. 13 TM behavioral model.

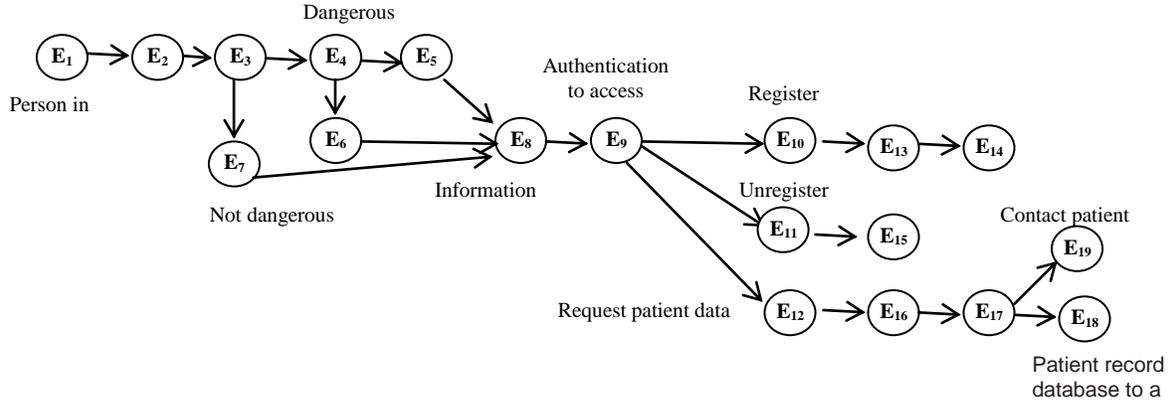

Fig. 14 Behavioral model of the Mentcare system.